\documentclass[
	sigconf,
 	screen,
 	nonacm,
]{acmart}

\usepackage{amsmath}
\usepackage{graphicx}
\usepackage{xcolor}
\usepackage{tabularx}
\usepackage{multirow}
\usepackage{hyperref}

\DeclareMathOperator*{\diag}{diag}
\DeclareMathOperator*{\dm}{diagM}
\DeclareMathOperator*{\argmin}{arg\,min}

\newcommand{\normsq}[1]{\left\lVert#1\right\rVert^2}
\newcommand{\frobnormsq}[1]{\normsq{#1}_F}

\newcommand{\R}{\mathbb{R}}

\newcommand{\citeauthorplain}[1]{\begin{NoHyper}\citeauthor{#1}\end{NoHyper}}

\AtBeginDocument{%
	\providecommand\BibTeX{{%
			\normalfont B\kern-0.5em{\scshape i\kern-0.25em b}\kern-0.8em\TeX}}}

\begin{document}

	\title{Modelling Users with Item Metadata for Explainable and Interactive Recommendation}

	\author{Joey De Pauw}
	\email{joey.depauw@uantwerpen.be}
	\orcid{0000-0002-1417-922X}
	\affiliation{%
	  \institution{University of Antwerp}
	  \streetaddress{Middelheimlaan 1}
	  \city{Antwerp}
	  \country{Belgium}
	}

	\author{Koen Ruymbeek}
	\email{koen.ruymbeek@uantwerpen.be}
	\orcid{0000-0003-0956-0779}
	\affiliation{%
		\institution{University of Antwerp}
		\streetaddress{Middelheimlaan 1}
		\city{Antwerp}
		\country{Belgium}
	}
	
	\author{Bart Goethals}
	\email{bart.goethals@uantwerpen.be}
	\orcid{0000-0001-9327-9554}
	\affiliation{%
		\institution{University of Antwerp}
		\streetaddress{Middelheimlaan 1}
		\city{Antwerp}
		\country{Belgium}
	}
	\affiliation{%
		\institution{Monash University}
		\city{Melbourne}
		\country{Australia}
	}
	
	\renewcommand{\shortauthors}{J. De Pauw et al.}

	\begin{abstract}

Recommender systems are used in many different applications and contexts, however their main goal can always be summarised as ``connecting relevant content to interested users''.
Personalized recommendation algorithms achieve this goal by first building a profile of the user, either implicitly or explicitly, and then matching items with this profile to find relevant content.
The more interpretable the profile and this ``matching function'' are, the easier it is to provide users with accurate and intuitive explanations, and also to let them interact with the system. 
Indeed, for a user to see what the system has already learned about her interests is of key importance
for her to provide feedback to the system and to guide it towards better understanding her preferences.

To this end, we propose a linear collaborative filtering recommendation model that builds user profiles within the domain of item metadata, which is arguably the most interpretable domain for end users. Our method is hence inherently transparent and explainable.
Moreover, since recommendations are computed as a linear function of item metadata and the interpretable user profile, our method seamlessly supports interactive recommendation.
In other words, users can directly tweak the weights of the learned profile for more fine-grained browsing and discovery of content based on their current interests.


We demonstrate the interactive aspect of this model in an online application for discovering cultural events in Belgium.
Additionally, the performance of the model is evaluated with offline experiments, both static and with simulated feedback, and compared to several state-of-the-art and state-of-practice baselines.

	\end{abstract}

	\ccsdesc[500]{Information systems~Search interfaces}
	\ccsdesc[300]{Information systems~Collaborative search}
	\ccsdesc[300]{Information systems~Personalization}
	\ccsdesc[300]{Information systems~Learning to rank}
	
	\keywords{interactive recommendation, personalization, transparency, explainability}

	\begin{teaserfigure}
		\centering
		\includegraphics[width=\textwidth]{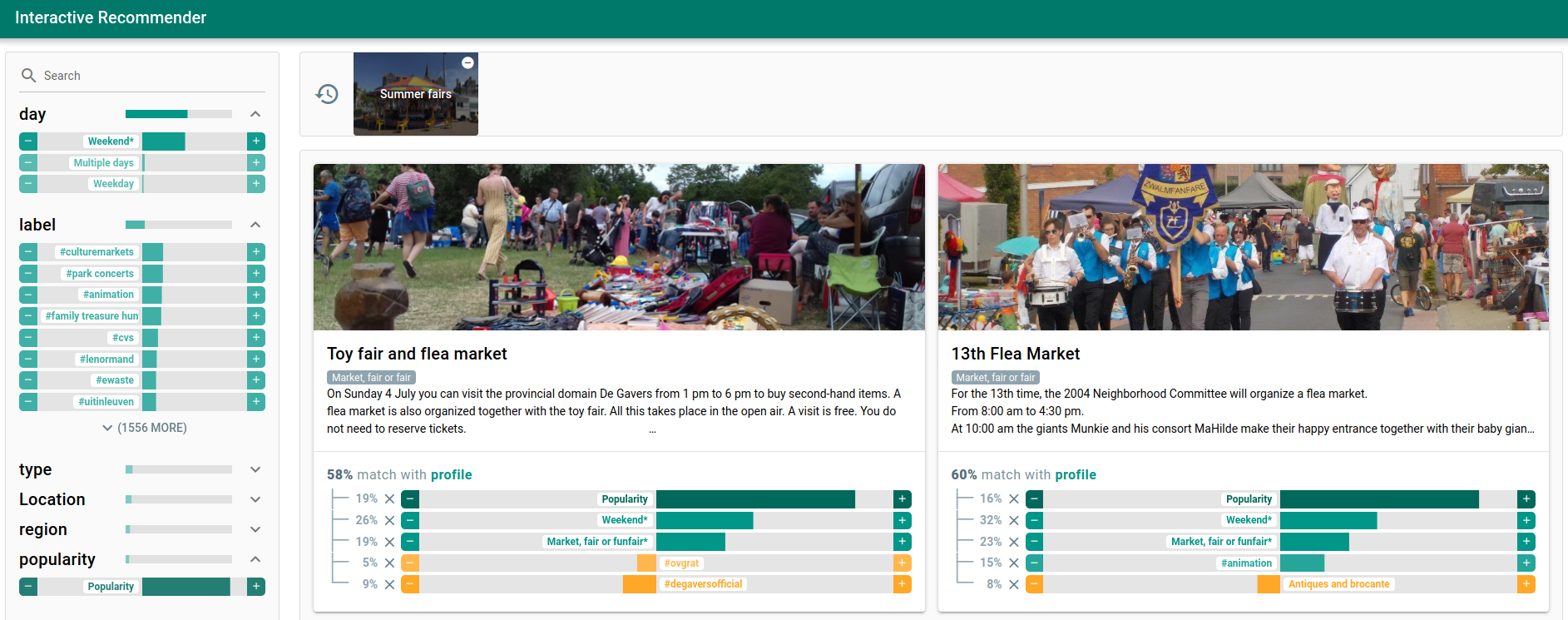}
		\caption{Screenshot of the interactive recommender system interface hosted on \url{https://tease-recommender.info}.}
		\Description{A screenshot of the application. To the left we see the user profile features and on the right hand side there are recommended events with accompanying explanations.}
		\label{fig:app}
	\end{teaserfigure}

	\maketitle


\section{Introduction}
\label{sec:introduction}
	Recommender systems are everywhere these days, whether a user notices it or not; from browsing videos on streaming sites to online advertisements to personalized news feeds on social media. The rapid increase of available content has made their task more important than ever.
	Within the field of recommender systems, two main approaches can be discerned: \emph{content-based filtering} and \emph{collaborative filtering}~\cite{adomavicius2005toward}. Where the former bases its recommendation on similarity defined by item metadata, the latter leverages user interaction data to compute recommendations. Typically we find that, because of their use of item metadata, content-based filtering methods are easier to make explainable and they can usually deal better with the problem of item cold start~\cite{burke2002hybrid, ramon2020metafeatures}. On the other hand, collaborative filtering often leads to higher ranking accuracy and overall better recommendations because they are more informed of user behaviour~\cite{burke2002hybrid}.
	
	Whichever recommendation method is used, it is found that a user's trust in the system can be improved with truthful and relevant explanations, as explanations provide both context for the recommendations and insight into the system~\cite{herlocker2000explaining}. Furthermore, explanation also help the users to accept the given recommendation, to find relevant content faster and to increase the overall ease of use of the system~\cite{herlocker2000explaining, zhang2018explainable, tintarev2015explaining, zhang2018exploring}. However, not all explanations and explanation types are equally informative and their usefulness also depends on the recommendation scenario and the current goals of the user~\cite{tran2021users, tintarev2015explaining}.
	
	For example, collaborative filtering algorithms typically have explanations based on the computed similarities between users and/or items~\cite{tintarev2015explaining}. Content based filtering approaches in contrast can also exploit item metadata for feature based explanations, which are found to be more interpretable due to their typically smaller domain that is easier to understand~\cite{ramon2020metafeatures}. However, with content based filtering algorithms comes the unfortunate drawback that they are less informed to compute the actual recommendations~\cite{burke2002hybrid}. Alternatively, explanation methods can be designed as post-processing steps~\cite{musto2019linked, papadimitriou2012generalized, peake2018explanation}. This allows them to be used in combination with any recommendation algorithm, however, with the caveat that the \emph{fidelity} is high enough. There is no guarantee that explanations computed this way actually reflect what the algorithm has learned and hence they are of limited use for gaining insight into the underlying model.%
	$\quad$\looseness=-1
	
	The notion of understanding the recommendation model itself is referred to as \emph{transparency} in this paper. In general, the more transparent your underlying model, the easier it is to provide accurate and interpretable explanations~\cite{molnar2020interpretable}. Furthermore, having a more transparent model also benefits the end user and the practitioner directly because it can help them understand the recommender system as a whole. So called ``black box models'' meanwhile suffer from a complete lack of transparency and are risky to deploy, tedious to debug and near impossible to trust by end users as their output cannot trivially be predicted or understood by humans~\cite{molnar2020interpretable}.
	
	To overcome the above mentioned limitations of explanations and to improve transparency in recommendation algorithms, we propose a hybrid recommender model~\cite{burke2002hybrid} called TEASER for ``Transparent \& Explainable Aspect Space Embedding Recommender''. It uses implicit feedback interaction data to learn the similarities between aspects of items. In other words, our method infers from the history of items that users consume, which aspects they might like, and subsequently recommends new items based on the learned profiles. As a result we combine the benefits of both collaborative and content based filtering explanations to achieve a fully transparent recommendation algorithm. 
	
	In essence, TEASER is a linear regression model that learns a single matrix to connect user histories to item metadata. Thanks to its simple formulation without complex interactions, transformations or non-linearities, we achieve many benefits: intuitive and accurate explanations of recommendations, a fully transparent and interpretable user profile and finally support for user feedback on the provided explanations, also called interactive recommendation. To the best of our knowledge, no hybrid linear recommendation model with a comparable degree of explainability, transparency and interactiveness exists in the literature.
	
	This paper is structured as follows. Section~\ref{sec:model} explains the proposed model TEASER. The transparent, explainable and interactive aspects of the model are subsequently discussed and demonstrated in Section~\ref{sec:application}. Section~\ref{sec:experiments} shows the results of offline experiments to assess the accuracy of TEASER, both in a static setting, and with simulated feedback. Finally, Section~\ref{sec:related_work} outlines the related work and we conclude the paper in Section~\ref{sec:conclusions}.


\section{Proposed Model}
\label{sec:model}

TEASER is a hybrid recommender model for implicit feedback data. It is similar in conception to the well-known linear regression model EASE~\cite{steck2019embarrassingly}, and in fact it is based on one of its variants, namely EDLAE~\cite{steck2020autoencoders}.
We adapt the training objective of EDLAE by replacing the decoder matrix such that the user embeddings ``take on the meaning'' of the item metadata features. This choice was motivated by previous work that demonstrates how using metadata for explanation can greatly increase the understandability for end users~\cite{ramon2020metafeatures}.
Indeed, though EASE and EDLAE are also linear models, one can argue that they are still not understandable due to the intractable scale of their features~\cite{lipton2018mythos, molnar2020interpretable}. Condensing the learned information down to less and more interpretable weights is the main design choice behind TEASER.

Section~\ref{sec:definition}, explains how TEASER works. Then Section~\ref{sec:model_properties} discusses the benefits of our approach in detail, such as why it is transparent and explainable, and how it can be made interactive when used in an application.

\subsection{Definition}
\label{sec:definition}

Let $X \in \{0, 1 \}^{m \times n}$ be the binary interaction matrix with $m$ the number of users and $n$ the number of items, then the EDLAE model is defined as
\begin{gather}
    \begin{aligned}
        \hat{E}, \hat{D} = \argmin_{E,D} & \frobnormsq{ X - X\left(ED^\top - \dm( \diag(E D^\top)\right)} + \\& 
        \lambda \frobnormsq{ED^\top - \dm( \diag(ED^\top)) }
    \end{aligned} \label{eq:edlae}
\end{gather}
with $\dm(.)$ the diagonal matrix with given diagonal, $E, D \in \R^{n \times k}$ the encoder respectively the decoder, and rank $k$ smaller than $\min(m, n)$. Notice that in this definition it is not possible for the model to learn anything from the diagonal elements.
By eliminating the diagonal of $ED^\top$ from the training objective, we prevent the model from overfitting towards the identity as it can no longer learn to predict items based on their own presence in the history~\cite{steck2020autoencoders}.

\begin{figure}
	\centering
	\includegraphics[width=\linewidth]{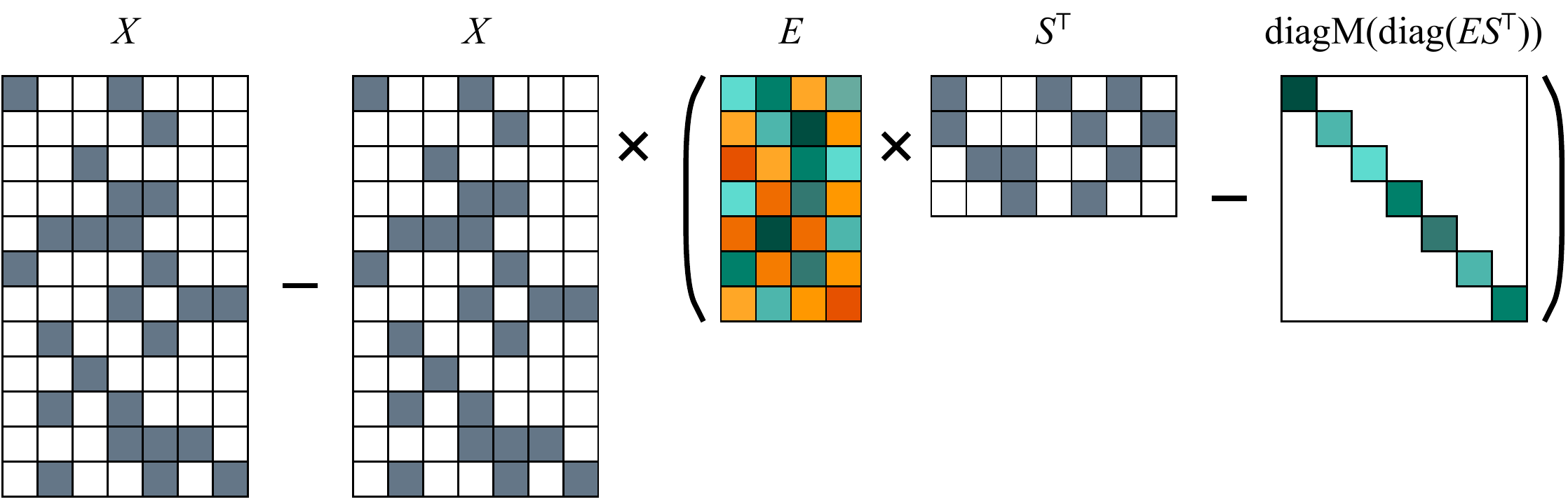}
	\caption{Graphical illustration of TEASER. ``$\times$''~represents matrix multiplication, teal colours indicate positive values, white and grey are zero and one respectively, and shades of orange are used for negative values. For a coloured version of the illustration, the reader is referred to the web version of this paper.}
	\Description{The matrices of the loss function represented in a graphical way with intuitive dimensions.}
	\label{fig:teaser_matrices}
\end{figure}

TEASER modifies EDLAE to use the fixed item metadata as item embeddings instead. Since users and items are embedded in a common latent space, this effectively makes it so the dimensions of this space correspond to the item features.
In practice, all the features are first made categorical where needed and are then one-hot encoded to be used as \emph{tags} of items. These binary vectors are combined in the matrix $S \in \{0, 1\}^{n \times t}$ with $t$ the number of tags. In TEASER we use $S$ as the decoder: $D = S$ and $k = t$, where $t$ is also lower than the number of users or items.
As a result, the encoder $E$ now connects the interaction matrix $X$ with the metadata matrix $S$, which means that our model is a hybrid of collaborative filtering (via $X$) and content-based filtering (via $S$). See Figure~\ref{fig:teaser_matrices} for a graphical illustration of the model.

As a side-effect of replacing $D$ with $S$, it also becomes necessary to regularize $E$. Otherwise the loss in $E$ is not always convex and the optimisation problem becomes harder to solve. Intuitively, adding squared $\ell_2$-norm regularization on $E$ pushes the model towards using smaller, more similar weights (in magnitude) as it penalizes extreme values more. Considering that we want $E$ to learn similarities that are generalized to some extent, this is actually desirable behaviour.

This final change brings us to the complete optimisation problem of TEASER:
\begin{gather}
	\begin{aligned}
		\hat{E} = \argmin_{E} &\frobnormsq{X - X(ES^\top - \dm(\diag(E S^\top )))} \\
		+ \lambda_1 &
		\frobnormsq{ES^\top - \dm(\diag(E S^\top))} 
		+ \lambda_2 \frobnormsq{E}.  
	\end{aligned} \label{eq:teaser}
\end{gather}
We solved this optimisation problem via the Alternating Directions Method of Multipliers \cite{boyd2011distributed, gabay1976dual, glowinski1975approximation}.
The computational complexity does not depend on the number of users if $X^\top X$ is precomputed and is cubic in the number of items. In practice it is about equally fast as EDLAE. More details on the exact formulas can be found in Appendix~\ref{sec:app_derivation}.

\subsection{Benefits}
\label{sec:model_properties}

The training objective of TEASER is quite restrictive: recommendation scores are no longer computed from pairwise item-item weights, but rather from tag-item weights. This seemingly simple change compared to EDLAE however has several benefits for \emph{explainability}, \emph{transparency}, \emph{interactiveness} and \emph{item cold start}, as explained in the next subsections.

\subsubsection{Explainability}
	The first and foremost benefit of using item metadata, is that explanations of recommendations can take on the form ``because you like \{outdoors activities\}'', or any other aspect of the item. These kinds of explanations can trivially be computed from the model, and are even the actual weights that lead to the recommendation. Namely, an item's score is calculated by summing over the aspects of that item in the ``user profile''. Given user history $\vec{x}$, an item $i$'s score is computed as $\langle\vec{x}E, S_{i}\rangle$. It is even possible to normalize this sum to compute the contribution of every aspect to the final score. We can hence give even more detailed \emph{item explanations} of the form ``Recommendation based for \{$37\%$\} on your affinity score of \{$0.4$\} with \{outdoors activities\}''. Naturally, the affinity score on itself is not very informative as it only gains meaning relative to other affinity scores.
	That is why they are scaled to fit in the range $[-1, 1]$ based on the minimum and maximum across all affinity scores of the user.
				
	Secondly, alongside item explanations, our method can also explain the \emph{profile} it has learned of the user. The aforementioned affinity scores with certain aspects are simply the user embedding ($\vec{x}E$) in our model. The higher the score, the more the system expects the user has interest in this aspect. Note that the user embedding is not based on just the aspects of the items in their history, as that would reduce our model to a trivial content based algorithm. Rather $E$ is learned from the interaction matrix as well as the metadata. Each row $E_i$ reflects which aspects a user is expected to find interesting (or not) in other items, if she \emph{consumed} item $i$.  For example when recommending events, if someone likes an event in Brussels, they can also be given a higher affinity with the neighbouring cities and villages or even with other big cities in Belgium, if this signal is present in the interaction data.
	
	
\subsubsection{Transparency}
	Furthermore, since the only learned weights are those in the encoder $E$ (see Figure~\ref{fig:teaser_matrices}), we make optimal use of the information that a user already knows (her history in $X$) or that she wants to discover (information about items in $D$).	Consider the example of recommending events: the user is most likely aware of her history of previously visited events and wants to find new interesting things to do. Typical explanations of the form ``Event A is recommended because you liked event B'' provide the user with some notion of similarity, however which similarity remains unknown. \textit{Are the two events similar, or are the events visited by similar users? In what ways are the events similar, or what do those similar users have in common with me?} In other words, the user profile is transparent, but the \emph{matching function} that computes recommendations based on the profile is not.
	
	By choosing our embeddings to be in the domain of item metadata, we effectively bridge this gap. The profile explanation gives new information to users about what the model has learned from them, while the item explanations tell users more about what is recommended to them and exactly for which aspects.	Note that information about the item is information that the user would want to know either way in order to assess whether the item is something for them. In conclusion, on top of a transparent profile, our \emph{matching function} (the decoder) is hence also interpretable because it is exactly the item metadata matrix.

\subsubsection{Interactiveness}
	The final benefit related to the transparency of our model lies in the fact that the explanations are exactly the weights of the model, and not the result of a post-processing step to approximate what the model really computed.	This means that, when a user gives feedback on the explanations, the feedback can seamlessly be integrated into the model and be used for interactive recommendation.
	
	Say for example that a user is less in the mood for outdoors activities today, but she finds that the recommender has based most of its suggestions on her recent participation in a series of marathons and open-air parties. Then she can simply decrease the weight of the \emph{outdoors tag} in her profile to uncover other relevant content for her.
	More formally, a user can increase or decrease the weight of specific aspects in her profile as follows. Denote by $\vec{f}$ the vector of user feedback for each aspect, then the formula for the user profile simply becomes: $\vec{x}E + \vec{f}$.
	
	In practice, we also introduce a scaling coefficient $c$ between zero and one to represent how \emph{certain} the model is of the user profile. This coefficient allows us to transform the affinity scores to the range $[-c, c]$ which has two advantages. First, choosing $c < 1$ allows the user to increase even the highest affinity score, which would otherwise already be at the maximal score of 1. Second, by letting this parameter depend on the user, we can tune how extreme the learned affinities are displayed. For example if a user only viewed one item, we might want to scale the overall affinity scores down compared to when they have a longer history to reflect the initial uncertainty due to lack of evidence.
	
	Naturally there are many ways to model the \emph{certainty} parameter, ranging from statistical modelling based on all user profiles to simply choosing a fixed value for all users. A simple yet intuitive heuristic based on the user history length $h$ is used in this work, where we scale up the certainty from 0.2 to 0.8 over history lengths from zero to three with the following function: $c=\textrm{min}(0.2 + 0.2h, 0.8)$. The choice of how to model certainty depends on the application and domain.

\subsubsection{Cold Start}
	A common problem in recommender systems is the cold start problem, i.e.\ what recommendations can you give to a user that has not interacted with the system yet (user cold start) and to whom would the system recommend an item if no user has ever interacted with this item (item cold start)? Content-based filtering recommenders do not suffer from item cold start, as recommendations are solely based on item metadata and not on interactions. This benefit carries over to our hybrid method since only the item's metadata needs to be added to $D$ for it to become recommendable.%
	$\quad$\looseness=-1
	
	To tackle the user cold start problem, we add \emph{popularity} as a feature of an item. An extra column is added to the matrix $D$, where every item gets a popularity score between zero and one. This score is computed as their total interaction count divided by that of the most popular item. Then, by having a cold start user start with some weight for the popularity tag, they will get the most popular items as initial recommendations.


\section{Application}
\label{sec:application}
	

	To demonstrate the interactive, transparent and explainable aspects of our model in action, we developed a publicly available demo that is hosted on \url{https://tease-recommender.info}. In this application, the user can interact with the TEASER model to browse for events or other things to do in Belgium. Out of the three datasets that are considered in this paper (see Section~\ref{sec:experiments}), we chose the proprietary dataset provided by Publiq for its rich and intuitive metadata. Every item also has at least a type, title, description and image which helps to display the events in a user-friendly way.
	
	A screenshot of the application is given in Figure~\ref{fig:app}. Note that the interface is written in Dutch due to the scope of the demo, but it can easily be translated into English by most modern browsers, as shown in the screenshot. The first thing a user sees when browsing to the website, is an interactive tour to guide her through the application. We summarise this tour here for completeness.
	
	On the left-hand side is a list of \emph{categories} and the respective \emph{tags} they contain. For example the category `day' has three tags: `Weekend', `Multiple days' and `Weekday'. For each tag the model computes an affinity score based on the user history. This collection of tags and scores is what we call the \emph{user profile}. Higher affinity scores are shaded more teal and the progress bar is filled more to the right, whereas for negative scores the bar is filled to the left and shaded in orange. Additionally, we can see the estimated impact that each category has on the recommendations with the positive-only progress bars next to them. These sum up to one across all features.$\quad$\looseness=-1
	
	On the top-right then, we find an ordered list of recently viewed events in the \emph{user history}. These are what make up the profile on the left and can help the user make the connection between the two. It is also possible for the user to remove any of the history items and receive recommendations based on the remaining ones. Though it is not inherently required by the proposed model, we included this feature more for experts to play around with the recommendations and have control over the entire state.
	
	Finally on the remainder of the page, we show personalized \emph{recommendations} for the user. Alongside the basic information of each event, we also include detailed information about why the event was selected specifically for the user. These \emph{item explanations} are computed as mentioned in Section~\ref{sec:model_properties} and only the top-5 explanations are selected to be displayed if they have a contribution of $5\%$ or more on the final score in absolute value. One can see that this also allows \emph{negative explanations} to be given, indicating that the model recommended an event to you, despite knowing that you might not like it for specific reasons. This of course is a design choice and can be disabled if negative explanations are undesirable, for example in other domains.
	
	To dive deeper into the explanations, we first dissect the overall \%-match score into its normalized sum of aspects. Here the percentages show how much each aspect contributed to the final score. The aspects themselves are the same as in the user profile. We hence provide a way for the user to both learn more about the event \emph{and} about why the model thinks she will like it, effectively ranking the most interesting information about the event for the user.
	
	The capstone feature of TEASER and the demo is its ability to interact with the user. As can be seen, each tag of the user profile also has a plus and a minus button. These behave as expected and increase respectively decrease the weight of the tag, which allows the user to manually indicate her preferences on top of what the model has learned.
	
	Naturally, the full list of tags is too large to manually sift through and luckily it is not necessary for a user to do so. The most influential tags are always listed first in the user profile, and since they are more likely to contribute to top-ranked items, they also appear as explanations for the events. Additionally, in the user cold start case it is much more efficient for the user to express interest in a few events and let the model estimate her affinity with each tag, than for her to manually set them herself.
		


\section{Experiments}
\label{sec:experiments}

	The main benefit of our model over classic recommendation algorithms is its transparency. Where the demo application was built to attest to how natural the explanations feel and to demonstrate the extent of the transparency of the model, we also perform several experiments to evaluate the capacity of the model to retrieve relevant content.
	
	Our novel TEASER model is compared on three different datasets with four baseline methods:
	\begin{itemize}
		\item EDLAE~\cite{steck2020autoencoders}, the unconstrained variants of TEASER,
		\item WMF~\cite{hu2008collaborative, pan2008one}, a well established matrix factorization model,
		\item EASE~\cite{steck2019embarrassingly} a state-of-the-art linear regression model by which EDLAE was inspired, and finally
		\item item k-nearest neigbours~\cite{deshpande2004item} which is a simple heuristic method that often works well in practice.
	\end{itemize}
	
	Additionally, we also compare TEASER with two other algorithms that share some of the benefits of explainability, transparency and interactiveness. The first is a trivial baseline we developed ourselves, namely a constrained variant of WMF we call ``WMF-S''. In WMF, scores are calculated based on a factorisation $U V^\top$ of the interaction matrix $X$. We choose $V = S$ and only optimise for the matrix $U$ in exactly the same way as WMF. This model needs more prediction time than TEASER (due to it needing to learn the user embedding again after each interaction) and it also has less interpretable weights (only per user weights, no global E).
	
	The second algorithm is a combination of TEASER and the baseline method EASE. We posit that the predictions of TEASER can be \emph{combined} with those of another recommender model, while maintaining the validity of the explanations. To achieve this, we essentially want to encode an \emph{`and'} relation between the two predictions, as this would mean that both models need to see the value in an item for it to be recommended. This is realized by taking the element-wise product between the predicted scores larger than zero (and setting the negative ones to zero), which corresponds to using the geometric mean of the predictions. Recommendations from this variant called TEASER $\odot$ EASE can still be explained in the same way, however these explanations now only reflect part of the logic.

	Section~\ref{sec:data} and \ref{sec:setup} concern the datasets and the experimental setup.
	The results and disussion can be found in Sections~\ref{sec:results} and~\ref{sec:sim_results}.

\subsection{Data}
\label{sec:data}

	In addition to the proprietary dataset, Publiq, that is used in the application, we also report experimental results on two publicly available datasets: The MovieLens20M~\cite{harper2015movielens} and Amazon Video Games~\cite{he2016ups} datasets. The number of users and items, the density and other statistics of the three datasets can be found in Table~\ref{tab:results} below the results. We chose these datasets for their wide variety of metadata (Publiq), to have one large scale and more dense dataset (ML20M), and one smaller and more sparse dataset (AVG).\looseness=-1
	
	As item metadata, we selected the most informative features that were provided with the datasets. For the ML20M dataset we also augmented the provided metadata with information from IMDb~\cite{imdb_data}. All features were converted to a one-hot encoding after filtering. 
	For more information we refer to Table~\ref{tab:results} and the source code\footnote{Source code available: \url{https://github.com/JoeyDP/TEASER}.}.\looseness=-1

\subsection{Experimental Setup}
\label{sec:setup}

	A strong generalization scheme with distinct train, validation and test users is employed in our experiments. First the user base is divided in three sets: a training set for learning the models, a validation set for optimising hyperparameters using grid search and finally a test set on which the final results are reported. Furthermore, the users in the validation and test sets are ensured to have at least 5 interactions, which are divided per user in a history (80\%) and ground truth (20\%) part. These parts are respectively used to generate recommendations and compute the reported metrics.
	
	Since the goal of our experiments is to evaluate how well the proposed model can retrieve relevant content given the transparency constraint, we chose three metrics\footnote{Metric definitions of~\cite{liang2018variational} are used.} that are representative of different tasks:
	\begin{itemize}
		\item recall@20, to evaluate the `standalone' performance.
		\item recall@100, to see whether the model can make a `short list' of relevant items, for example to be reranked.
		\item nDCG@100, a metric that also takes the rank of items into account, as a global indicator of the ranking accuracy (also the target of the grid search).
	\end{itemize}
	
	For a more fair comparison, some hyperparameters were chosen outside of the grid search. Namely the number of factors of EDLAE was fixed to the number of tags used in TEASER and the hyperparameters of TEASER $\odot$ EASE were not optimised again but rather taken from their respective optima. Note that for WMF we did not use the number of tags as the number of factors because this method does not scale well with the number of factors.

	In addition to the typical static evaluation, we also conducted experiments with \emph{simulated feedback} for the interactive models. By simulating feedback, we can estimate the improvement a recommender system is expected to gain by interacting with the user. Of course this is not a perfect substitute for a real user study because numerous assumptions have to be made, among which the feedback sampling strategy and the cross-validation method~\cite{konstan2012recommender}.
	Furthermore, there are also known biases in offline evaluation in general, most notably the missing-not-at-random or feedback loop phenomenon where some positives are more likely to be missing due to the policy under which the data was recorded~\cite{canamares2020offline, steck2010training, marlin2009collaborative}. Nevertheless, the results are still representative of the task they model, and can be used as a proxy for the expected effect of users interacting with the system.
	
	We propose two interactive scenarios: in the first scenario each test user gives a positive boost to one tag of the items in their ground truth set, and in the second scenario two tags are sampled.
	Note that a similar experimental setup is often used to evaluate critiquing methods where a user's critique on an item is sampled from the ground truth set and the amount of iterations until the item appears in the top-K is reported~\cite{luo2020latent, li2020ranking, antognini2021fast}. In our experimental setup however we can have multiple target items in the test set, so we report the improvement in ranking accuracy instead.
	
	Our simulation is set-up such that the strength of the positive signal is three out of five, which represents three clicks and raises the affinity by a little over half. Considering that the values are scaled based on the highest tag score and the certainty function that is capped at 0.8, this can essentially boost a tag originally at 0 to a little below the highest learned tag. Choosing the middle ground of 3 for signal strength is of course an arbitrary choice and the improvements can be expected to be less or more extreme given a different choice. Selecting which tag to boost is also based on a heuristic, namely by sampling from all the tags of items in the test set, weighted by occurrence count. As such, the simulation introduces randomness to the evaluation. In order to reduce the effect of randomly picking a very good (or very bad) tag, we take the average of three simulation runs for each user, which is found to give stable results for all experiments.
	

\subsection{Static Results}
\label{sec:results}

	\newcommand*{\best}[1]{\textbf{#1}}
	\newcommand*{\improv}[1]{+#1}

	\begin{table*}
		\caption{Static experimental evaluation of the baseline and proposed algorithms on three datasets. Best result per metric is typeset in bold. Properties of the datasets are listed below the results.}
		\label{tab:results}
		\begin{tabular}{lrrrrrrrrrrrr}
			&& \multicolumn{3}{c}{\textbf{MovieLens20M}} && \multicolumn{3}{c}{\textbf{Amazon Video Games}} && \multicolumn{3}{c}{\textbf{Publiq}} \\
			\cmidrule{3-5} \cmidrule{7-9} \cmidrule{11-13}
			&& Recall & Recall & nDCG && Recall & Recall & nDCG && Recall & Recall & nDCG \\
			\textbf{models} && @20 & @100 & @100 && @20 & @100 & @100 && @20 & @100 & @100 \\
			\cmidrule{1-1} \cmidrule{3-5} \cmidrule{7-9} \cmidrule{11-13}
			\textbf{static baselines:} && & & && & & && & & \\
			EASE && \best{0.395} & \best{0.637} & \best{0.422} && 0.218 & 0.369 & 0.145 && 0.481 & 0.639 & 0.329 \\
			EDLAE && 0.389 & 0.635 & 0.414 && 0.209 & 0.372 & 0.143 && 0.488 & 0.657 & 0.322 \\
			ItemKNN && 0.310 & 0.505 & 0.328 && 0.150 & 0.309 & 0.103 && 0.434 & 0.614 & 0.299 \\
			WMF && 0.382 & 0.626 & 0.406 && \best{0.226} & \best{0.433} & \best{0.157} && \best{0.524} & \best{0.704} & \best{0.337} \\
			\cmidrule{1-1} \cmidrule{3-5} \cmidrule{7-9} \cmidrule{11-13}
			\textbf{interactive models:} && & & && & & && & & \\
			TEASER && 0.110 & 0.233 & 0.136 && 0.130 & 0.280 & 0.095 && 0.348 & 0.551 & 0.236 \\
			WMF-S && 0.225 & 0.420 & 0.266 && 0.041 & 0.114 & 0.034 && 0.290 & 0.514 & 0.190 \\
			TEASER $\odot$ EASE && 0.356 & 0.601 & 0.387 && 0.189 & 0.366 & 0.132 && 0.495 & 0.642 & 0.326 \\
			\midrule
			\multicolumn{1}{l}{\textbf{dataset prop.:}} && & & && & & && & & \\
			\# users && \multicolumn{3}{c}{138~287} && \multicolumn{3}{c}{24~072} && \multicolumn{3}{c}{46~075} \\
			\# items && \multicolumn{3}{c}{20~720} && \multicolumn{3}{c}{10~622} && \multicolumn{3}{c}{15~000} \\
			inter. density && \multicolumn{3}{c}{0.349\%} && \multicolumn{3}{c}{0.068\%} && \multicolumn{3}{c}{0.019\%} \\
			\# tags && \multicolumn{3}{c}{3~504} && \multicolumn{3}{c}{1~688} && \multicolumn{3}{c}{2~158} \\
			tag density && \multicolumn{3}{c}{0.288\%} && \multicolumn{3}{c}{3.381\%} && \multicolumn{3}{c}{0.578\%} \\
			features && \multicolumn{3}{m{3cm}}{year, runtime, genres, tags, directors, writers} && \multicolumn{3}{m{3cm}}{price, brand, categories, keywords} && \multicolumn{3}{m{3cm}}{type, audience, theme, labels, price, region, \ldots} \\
			\bottomrule
		\end{tabular}
	\end{table*}

	The results of the static experiments are listed in Table~\ref{tab:results} where the 
	four baselines that do not use item metadata are listed in the first four rows, followed by the proposed transparent and explainable models. A first general trend can be observed between the more explainable (and constrained) models and the baselines: the predictive accuracy of the former class is always lower than that of the best performing unconstrained baseline in a static setting. This is an expected and known trade-off when constraining machine learning models to be more transparent~\cite{molnar2020interpretable}.
	
	In this case the accuracy of recommendations and the quality of explanations depend heavily on the amount and quality of the item metadata, which is a known issue in tag-based systems~\cite{golder2006usage, sen2007quest, vig2009tagsplanations}. For example if two distinct items have the same (incomplete) set of features, there is no way for our algorithms to distinguish them. Similarly, redundant tags that bear the same meaning can skew the tag importance and lead to less user-friendly explanations. This however is a data quality problem outside of our control.
	
	
	It should also be noted that quantity cannot compensate for quality in this setting. Even if a lot more information is available about the items, if this information is not discriminative for predicting user interests, the models still will not be able to benefit from it. As an example, consider the large difference in \emph{tag density} between the Amazon Video Games (3.38\%) and the Publiq (0.58\%) datasets. Despite the much higher tag density in Amazon Video Games (which mostly comes from keyword mining), the performance drop in nDCG between the best baseline and TEASER on this dataset is still higher than that on the Publiq dataset. 
	
	A second observation that can be made is that, on top of the benefit of more explainable weights and faster predictions, TEASER is also more accurate than WMF-S on the Amazon Video Games and Publiq datasets. On the MovieLens dataset however we see the opposite. A possible explanation for this difference is that WMF-S is less constrained in what it can learn. Indeed, the user factors in WMF-S do not need to be a linear combination of item-factors as is the case for TEASER. We hypothesize that this additional freedom together with the fact that the MovieLens dataset has more dense interactions and less abundant and informative metadata, has lead to the observed results.

	
	Thirdly and finally, the results show that by combining an aspect based method with a well performing unconstrained baseline, we can effectively approach the best of both worlds. For all three datasets we find that TEASER $\odot$ EASE achieves a ranking accuracy between those of the respective individual models. This shows that, despite taking the product of scores, the models can still work together to obtain a better ranking while preserving the explainability, and only sacrificing a bit of transparency.
	Note that this specific combination of models was chosen to evaluate the validity of taking the product of scores and not to imply superiority of TEASER. The other combinations of models can be expected to enjoy similar benefits.\looseness=-1
	
\subsection{Simulated Feedback Results}
\label{sec:sim_results}
	
	\begin{table*}
		\caption{Simulated experimental results of the interactive models. Best result per metric per scenario (block) is typeset in bold and for nDCG the relative improvement to the static version of each model is included. First block is repeated from Table~\ref{tab:results}. TEASER $\odot$ EASE is abbreviated T $\odot$ E.}
		\label{tab:results_simulated}
		\begin{tabular}{lrrrrrrrrrrrrrrr}
			&& \multicolumn{4}{c}{\textbf{MovieLens20M}} && \multicolumn{4}{c}{\textbf{Amazon Video Games}} && \multicolumn{4}{c}{\textbf{Publiq}} \\
			\cmidrule{3-6} \cmidrule{8-11} \cmidrule{13-16}
			&& Recall & Recall & nDCG & impr. && Recall & Recall & nDCG & impr. && Recall & Recall & nDCG & impr. \\
			\textbf{models} && @20 & @100 & @100 & in \% && @20 & @100 & @100 & in \% && @20 & @100 & @100 & in \% \\
			\cmidrule{1-1} \cmidrule{3-6} \cmidrule{8-11} \cmidrule{13-16}
			\textbf{no sim.} && & & &&& & & &&& & && \\
			TEASER && 0.110 & 0.233 & 0.136 &&& 0.130 & 0.280 & 0.095 &&& 0.348 & 0.551 & 0.236 & \\
			WMF-S && 0.225 & 0.420 & 0.266 &&& 0.041 & 0.114 & 0.034 &&& 0.290 & 0.514 & 0.190 & \\
			T $\odot$ E && \best{0.356} & \best{0.601} & \best{0.387} &&& \best{0.189} & \best{0.366} & \best{0.132} &&& \best{0.495} & \best{0.642} & \best{0.326} & \\
			\cmidrule{1-1} \cmidrule{3-6} \cmidrule{8-11} \cmidrule{13-16}
			\textbf{one tag +3:} && & & &&& & & &&& & && \\
			TEASER && 0.128 & 0.267 & 0.152 & \improv{11.8} && 0.281 & \best{0.522} & 0.185 & \improv{94.7} && 0.475 & 0.715 & 0.328 & \improv{39.0} \\
			WMF-S && 0.239 & 0.442 & 0.279 & \improv{4.9} && 0.153 & 0.338 & 0.104 & \improv{205.9} && 0.394 & 0.629 & 0.261 & \improv{37.4} \\
			T $\odot$ E && \best{0.365} & \best{0.610} & \best{0.394} & \improv{1.8} && \best{0.307} & 0.499 & \best{0.198} & \improv{50.0} && \best{0.541} & \best{0.665} & \best{0.356} & \improv{9.2} \\
			\cmidrule{1-1} \cmidrule{3-6} \cmidrule{8-11} \cmidrule{13-16}
			\textbf{two tags +3:} && & & &&& & & &&& & && \\
			TEASER && 0.151 & 0.305 & 0.171 & \improv{25.7} && \best{0.478} & \best{0.714} & \best{0.294} & \improv{209.5} && \best{0.625} & \best{0.858} & \best{0.423} & \improv{79.2} \\
			WMF-S && 0.256 & 0.467 & 0.293 & \improv{10.2} && 0.357 & 0.621 & 0.224 & \improv{558.8} && 0.500 & 0.732 & 0.336 & \improv{76.8} \\
			T $\odot$ E && \best{0.373} & \best{0.617} & \best{0.400} & \improv{3.4} && 0.353 & 0.525 & 0.224 & \improv{69.7} && 0.566 & 0.671 & 0.375 & \improv{15.0} \\
			\bottomrule
		\end{tabular}
	\end{table*}

	A second set of experiments with simulated user feedback was conducted and its results are displayed in Table~\ref{tab:results_simulated}. The first observation we can make is that there is again a large difference in performance and performance gain between the MovieLens dataset and the other two datasets. Despite the relatively low scores (and large room for improvement) of the interactive models TEASER and WMF-S on MovieLens compared to the best baseline, it is clear that the simulated feedback did not have a big impact. This confirms our intuition that the metadata of the MovieLens dataset is not abundant and discriminative enough to train the proposed models. If this signal were stronger, we would expect to see similar improvements as with the other two datasets. As this is not the case, we can assume that the learned profiles in combination with the item features are not informed enough to provide good recommendation, even with feedback of the user.
	
	The other two datasets agree quite well with each other. Firstly, we find that even by adding positive feedback to only one tag for each user, the best interactive model TEASER $\odot$ EASE already surpasses the best baseline of Table~\ref{tab:results} for both datasets. Of course this is not a fair comparison from which we can conclude that one method is better than the other (as the static baselines are not given information about the ground truth items). However it does demonstrate the power of interactive recommendation. It is only because the proposed methods can take user feedback on tags into account that they are able to gain an advantage in this scenario.
	
	Secondly, if we look at the scenario with two boosted tags per user, we find that TEASER outperforms the ensemble method TEASER $\odot$ EASE by a large margin. In general, the ensemble method is expected to gain less benefit from feedback than the purely interactive methods as it balances the part that can take feedback into account with the static part. 
	In the limit, as more feedback is given, the interactive part starts to outperform the static part, at which point the combination is held back by the static part.
	The simulation results confirm this behaviour. Even when we take the higher base accuracy of TEASER $\odot$ EASE into account, we can see that the absolute improvement of the metrics is always bigger for TEASER and WMF-S than for the ensemble method.
	
	Thirdly and finally, we find that on the Amazon Video Games and Publiq datasets, the more sophisticated TEASER model also outperforms WMF-S like in the static experiments.


\section{Related Work}
\label{sec:related_work}

\paragraph*{Interactive Recommendation}
	Within the field of interactive recommendation, \citeauthorplain{he2016interactive} provide a framework for analyzing and comparing interactive recommender systems~\cite{he2016interactive}. They then apply this framework to 24 previously published systems which they first cluster by the primary objectives in their framework: \emph{transparency}, \emph{justiﬁcation}, \emph{controllability}, \emph{diversity}, \emph{cold start} and \emph{context}.
	
	TasteWeights~\cite{bostandjiev2012tasteweights} by \citeauthorplain{bostandjiev2012tasteweights} is perhaps the most similar interactive recommender system to our method. In their interactive music recommender, the profile is the user history and recommendations are based on a weighted combination of the recommendation scores computed from different sources like Wikipedia (content-based filtering) and Facebook (collaborative filtering). Interaction comes from allowing users to modify their history, the learned intermediate representation by the algorithms and the weights given to each algorithm or source. The main difference then, lies in the fact that we propose a single hybrid recommendation algorithm, where the learned representation is both based on item metadata (content based) and the interaction data (collaborative filtering).
	
	Other similar interactive recommender system include Tagsplanation~\cite{vig2009tagsplanations} and TagMF~\cite{loepp2019interactive}. These systems explain recommendations based on the estimated tag relevance for items combined with the computed tag preference of users. As such, they differ from our method in that we use binary tags that are generated from the item metadata.

\paragraph*{Explanations in Recommenders}
	Closely related to the field of interactive recommendation, is the topic of recommendation explanation and transparency that has gained much traction in recent years. We highlight three related works that have noteworthy overlap with the proposed method.
	
	First, \citeauthorplain{balog2019transparent} describe a set-based recommender with natural language explanations of the learned user profile~\cite{balog2019transparent}. Their method reaches full transparency by only computing the recommendation based on the top-k explanations that are actually presented to the user. Furthermore, their method also supports interactiveness or ``scrutability'' by allowing users to remove statements about their profile that they disagree with. Besides the use of item priors, they describe their method as content-based and do not make use of interaction data to compute recommendations.
	
	Second, aspect-based matrix factorization (AMF)~\cite{hou2019explainable} is proposed by \citeauthorplain{hou2019explainable}. AMF is a variation of matrix factorization that \emph{fuses} aspect information from review texts into the objective. This method is similar in that it models users as their affinity with aspects, however the notion of an aspect does not coincide with our use of item metadata. Most prominently, aspects are based on review texts and are learned both for users and for items, where we assume fixed item tags.
	
	Third, \citeauthorplain{graus2018let}~\cite{graus2018let} propose to not only explain the recommended items, but also the computed user profile. They illustrate the benefits this can bring for content-based recommenders, where the user profile is a collection of content features. Our work applies this principle in the context of collaborative filtering with a hybrid method.

\paragraph*{Tag-based Recommenders}
	A third line of related work to discuss is that of tag-based recommenders. In this setting, users are able to tag items and recommendations are based on the inferred preference of a user for (a subset of) tags~\cite{sen2009tagommenders, loepp2019interactive}. As such, the required data differs from our approach since we define tags as the one-hot encoding of item metadata, and hence inherent to the item and not as a function of users and items. The way recommendations are generated from this data is however very similar. Once the tag preferences of a user are inferred, recommendations are typically based on some similarity measure between those preferences and the tags of items~\cite{sen2009tagommenders}.

\paragraph*{Embedding Models with Item Features}
	The idea of using item features to constrain embeddings is certainly not new and has been presented in numerous forms. \citeauthorplain{vasile2016meta} for example factorise an item matrix extended with metadata~\cite{vasile2016meta}. Other works approach the problem by modelling item, content or user factors with shared information using various methods~\cite{yang2011like, fang2011matrix, agarwal2009regression, hou2019explainable}. The main distinction between our method and previous work is that TEASER does not constrain its embeddings to be close to the item metadata or to share factors with the content embedding.
	In TEASER, the item metadata matrix \emph{is} the embedding of the items, which gives our method the benefit of additional interpretability.

\paragraph*{Critiquing}
	Critiquing was first applied to recommender systems (under the name of item browsing) by \citeauthorplain{burke1996knowledge}~\cite{burke1996knowledge}. They developed a system where users start from an initial item and subsequently provide feedback (critique) on properties of the proposed items. For example: \textit{make it cheaper}, \textit{make it bigger} or \textit{make it faster}. More recently, the idea of critiquing recommendations and specifically critiquing the explanations of recommendations has grown into a popular direction of research~\cite{luo2020latent, li2020ranking, antognini2021fast}. The distinction between the proposed method and critiquing methods is twofold: first, TEASER provides user-centric explanations through the user profiles (feedback is not relative to an item) and second, the user is given insight into and control over how the recommendations are generated due to its simplicity and transparency.



\section{Conclusions and Future Work}
\label{sec:conclusions}

	A new highly transparent, explainable and interactive hybrid recommender, TEASER, is presented and evaluated. The main benefit of TEASER lies in the use of item metadata to build a profile of the user, which is then used to compute recommendations and their accompanying explanations. As such, the model is fully transparent and explainable with item metadata. This domain both scales well and is intuitive for the end user. Additionally, practitioners can also gain insight into their trained models as the learned representations reflect similarities that are discovered from the interaction data.%
	$\quad$\looseness=-1
	
	Furthermore, TEASER seamlessly enables interactive recommendation, where the user can provide feedback on the explanations and on the learned profile for the model to incorporate. We demonstrate the explainable and interactive aspects in an online web application and evaluate the ranking accuracy with offline static and simulated experiments. The experimental results show that, depending on the quality of the used metadata, TEASER is still able to provide decent recommendations, especially when combined with an unconstrained baseline. On top of that, when simulated feedback is included, we find that the interactive aspect of TEASER enables it to outperform its static counterparts with ease.
	
	\subsection{Future Work}
	\label{sec:future_work}
	
	For future work we consider several interesting extensions to the base model, among which:
	\begin{itemize}
		\item Learning the decoder matrix $D$ according to the sparsity pattern of S. This makes the model more flexible than being constrained to only binary entries. Essentially it would allow the model to learn that some features matter less (or more) for specific items. In the example of an event recommender, one can imagine that for a concert of an international artist, the location matters less than for a local party, as people will probably be willing to travel further for the former.
		\item Adding content based information to the learned similarities. By adding the following weighted extra term to the loss: $\delta\frobnormsq{S^\top - S^\top (ED^\top - \dm(\beta))}$ (which is trivial to optimise for), one can include simple content based matching into the objective. Previous work has demonstrated that this can improve recommendation accuracy, mostly when only very sparse interaction data is available~\cite{jeunen2020closed}.
		\item In addition to learning from item metadata, it is also possible to learn from user information. For example, by extending the learned user embedding ($\vec{x}E$) with user specific features like their age or other information that they might want their recommendations to be based on, we can add learnable weights in $D$ to take them into account.
		\item Though the number of features is already smaller than the number of items and hence more tractable to interpret, additional sparsity in $E$ may be desired. Adding a sparsity constraint like $\ell_1$-norm regularization on $E$ could lead to even more explainable solutions.
	\end{itemize}
	
	Other future work could be defined in the area of evaluation and in improving the user experience. Firstly, we argue that the explanations are intuitive and benefit the end user, but we do not have the results of a user study to establish this. Secondly, more effort can be put into the user cold start problem, for example by leveraging active learning to quickly gain the required information to build an accurate user profile. Thirdly, TEASER has a strong bias towards recommending items with similar metadata. This arises inherently from the imposed constraint unfortunately, but can easily be mitigated by including serendipity and diversity objectives in the prediction phase. We believe the ground work for including these objectives is already there, since the model is aware of item metadata and can provide a user profile based on these features out of the box.\looseness=-1

	\begin{acks}
		This work was supported by the \grantsponsor{fwo}{Research Foundation --- Flanders (FWO)}{https://www.fwo.be/} [\grantnum{fwo}{11E5921N} to J. De Pauw] and the 
		\grantsponsor{gov}{Flemish Government}{https://airesearchflanders.be/} under the \grantnum[https://airesearchflanders.be/]{gov}{``Onderzoeksprogramma Artificiële Intelligentie (AI) Vlaanderen''} programme.
		Special thanks to Noah Dani\"els for designing and developing the web application.
	\end{acks}

	\bibliographystyle{ACM-Reference-Format}
	\bibliography{references}
	
	\appendix

\section{Derivation of TEASER}
\label{sec:app_derivation}

Here we give a detailed derivation how the optimisation problem \eqref{eq:teaser} is solved. For this, the method of Lagrangian multipliers together with the Alternating Directions Method of Multipliers (ADMM) is used. To this end, we introduce an extra variable $\beta$ and solve the equivalent problem
\begin{gather}
	\begin{aligned}
		\hat{E} = \argmin_{E} &\frobnormsq{X + X \dm(\beta) - X ED^\top} \\ 
		+~\lambda_1 &
		\frobnormsq{ED^\top - \dm(\beta)} + \lambda_2 \frobnormsq{E}  
	\end{aligned} \label{eq:teaser_beta} \\
	\begin{aligned}
		\textrm{s.t.}&& \diag(E D^\top) & = \beta 
	\end{aligned} \label{eq:constr1}
\end{gather}

Concerning the constraint \eqref{eq:constr1}, we can write the augmented Lagrangian:
\begin{align}
		L(E, \beta, \gamma) =  &\frobnormsq{X + X \dm(\beta) - X ED^\top} \nonumber \\
		& + \lambda_1 \frobnormsq{ED^\top - \dm(\beta) }
 		 + \lambda_2  \frobnormsq{E} \nonumber \\
 		& + 2 \rho \gamma^\top (\beta - \diag(E D^\top)) + \rho \normsq{\beta - \diag(E D^\top) }_2  \label{eq:constr1_lagr}
\end{align}
where \eqref{eq:constr1_lagr} deals with constraint \eqref{eq:constr1}. The variable $\gamma$ represents Lagrange multipliers for the constraint. The scalar $\rho  > 0$ is a penalty parameter in the augmented Lagrangian and can be considered as a training-hyperparameter. Before explaining the formulas in detail, we want to point out that the term $\rho \normsq{\beta - \diag(E D^\top) }_2$ makes the optimisation problem hard to solve. We change this term in \eqref{eq:constr1_lagr} into
\begin{align*}
    \rho \normsq{\beta - \diag(E D^\top) }_2
    = & \rho \frobnormsq{E D^\top - \dm(\beta) }  \\
    &-\rho \frobnormsq{ E D^\top - \dm( \diag( E D^\top))} \\
    \approx & \rho \frobnormsq{E D^\top - \dm( \beta) } - \rho \frobnormsq{E D^\top} + \rho \normsq{\beta}_2.
\end{align*}

During the optimisation, $\beta$ will come closer to $\diag( E D^\top)$ and ADMM also converges under inexact minimization \cite{boyd2011distributed} if it becomes more and more exact  over the iterations. This means we are allowed to, instead of $L(E, \beta, \gamma)$, optimise for the loss function 
\begin{align}
		\tilde{L}(E, \beta, \gamma) =  &\frobnormsq{X + X \dm(\beta) - X ED^\top} + \lambda_1 \frobnormsq{ED^\top - \dm(\beta) } \nonumber \\&
 		 + \lambda_2  \frobnormsq{E} + 2 \rho \gamma^\top \left(\beta - \diag(E D^\top) \right) \nonumber \\
 		& + \rho \left( \frobnormsq{\beta - \diag(E D^\top) } - \frobnormsq{E D^\top} + \normsq{\beta}_2  \right).  
\end{align}

ADMM is an iterative method.  At iteration $k+1$, following scheme is used:
\begin{align}
    E^{(k+1)} & = \argmin_E \tilde{L} \left( E, \beta^{(k)}, \gamma^{(k)} \right) \label{eq:optim_E} \\
    \beta^{(k+1)} & = \argmin_\beta \tilde{L} \left( E^{(k+1)}, \beta, \gamma^{(k)} \right) \label{eq:optim_beta} \\
    \gamma^{(k+1)} & = \gamma^{(k)} + \diag\left( E^{(k+1)} D^\top \right) - \beta^{(k+1) }. \nonumber
\end{align}
We optimise \eqref{eq:optim_E} resp. \eqref{eq:optim_beta} by setting the partial derivation of $\tilde{L}$ towards $E$ resp. $\beta$ to zero and solve it. For solving \eqref{eq:optim_beta}, we obtain the following analytic formula 
\begin{equation} \label{eq:update_beta}  
\beta^{(k+1)} = \dfrac{ \diag(X^T X (- I  + ED^\top ))  - \rho \gamma + (\lambda_1 + \rho) \diag( E D^\top) }{ \diag(X^\top X + (\lambda_1 + 2 \rho) I ) }.
\end{equation}
where we omit the superscripts in $E$ and $\gamma$ for ease of notation.
The solving of \eqref{eq:optim_E} is more difficult as following equation needs to be solved for $E$,
\begin{equation} \label{eq:sylvester_E}
\begin{split} 
    &\left( X^\top X  + \lambda_1 I \right) E D^\top D  + \lambda_2 E = \\&
    \left( X^\top X \dm \left( 1 + \beta \right) + \rho \dm(\gamma + \beta + \lambda_1 \dm( \beta) \right) D
\end{split}
\end{equation}
where we omit the superscript in $\gamma$ and $\beta$.
We see that only the right-hand side of this equation changes in every iteration.
The equation \eqref{eq:sylvester_E} can be rewritten as a Sylvester equation. We do not explicitly rewrite it as a Sylvester equation but we apply the same ideas as in the Bartels-Stewart algorithm \cite{bartels1972solution}.
First, we make an eigendecomposition $U \diag(\mu) U^\top = X^T X$ and $V \diag(\eta) V^\top = D^\top D$, so that \eqref{eq:sylvester_E} is equivalent with
\begin{align}
    (\diag(\mu) + \lambda_1) Y^{(k+1)} \diag(\eta) + \lambda_2 Y^{(k+1)} & = F^{(k+1)} \label{eq:E_help}\\
    E^{(k+1)} & = U Y^{(k+1)} V^\top \nonumber 
\end{align}
with $$ F^{(k+1)} = U^\top \left( X^\top X \dm( 1 + \beta) + \rho \dm(\gamma + \beta) + \lambda_1 \dm( \beta) \right) D V$$
where also here we omit the superscripts in $\gamma$ and $\beta$. By solving \eqref{eq:E_help} column by column, it can be seen that following equations lead to the solution
\begin{align*}
    Y^{(k+1)} = F^{(k+1)} \odot G \qquad\qquad
    E^{(k+1)} = U Y^{(k+1)} V^\top
\end{align*}
with $G(i,j) = \left( \eta_j (\mu_i + \lambda_1) + \lambda_2 \right)^{-1}, i = 1, \hdots,m, j = 1, \hdots, t.$
We agree that taking the decomposition of $X^\top X$ is expensive, however it only needs to be computed once and can be reused in all iterations.

\end{document}